# Approaches to link RNA secondary structures with splicing regulation

**Mireya Plass[1] and Eduardo Eyras[2,3,*]**

[1] The Bioinformatics Centre, Department of Biology, University of Copenhagen, Ole Maaløes Vej 5, 2200 Copenhagen N, Denmark

[2] Computational Genomics, Universitat Pompeu Fabra, Dr. Aiguader 88, 08002, Barcelona, Spain

[3] Catalan Institution for Research and Advanced Studies (ICREA), Passeig Lluís Companys 23, 08010 Barcelona, Spain

* Corresponding author

## Summary

In higher eukaryotes, alternative splicing is usually regulated by protein factors, which bind to the pre-mRNA and affect the recognition of splicing signals. There is recent evidence that the secondary structure of the pre-mRNA may also play an important role in this process, either by facilitating or by hindering the interaction with factors and small nuclear ribonucleoproteins (snRNPs) that regulate splicing. Moreover, the secondary structure could play a fundamental role in the splicing of yeast species, which lack many of the regulatory splicing factors present in metazoans. This review describes the steps in the analysis of the secondary structure of the pre-mRNA and its possible relation to splicing. As a working example, we use the case of yeast and the problem of the recognition of the 3' splice site (3'ss).

# 1. Introduction

Splicing is the mechanism by which introns are removed from the pre-mRNA to create the mature transcript. In higher eukaryotes this process involves, apart from the core machinery of the spliceosome, many auxiliary factors, e.g. SR proteins or hnRNPs, which can enhance or block the recognition of splicing signals *(1)*. These factors allow the modulation of the splicing reaction and thus, the existence of alternative splicing.

During transcription, the synthesized RNA can fold *(2)*. Accordingly, secondary structures adopted by the pre-mRNA may influence splicing regulation. RNA structures can hinder the recognition of splicing signals by occluding them and preventing their recognition by spliceosome components. Alternatively, they could expose signals necessary for regulation. Interestingly, predicted secondary structures have been identified to aid the computational prediction of splice sites *(3, 4)* and genome-wide analyses have shown that conserved RNA secondary structures overlapping splice sites are related to alternative splicing *(5)*. Besides, these pre-mRNA structures can facilitate the recognition of splicing signals by shortening the distance between them *(6, 7)*. In other cases, RNA structures can regulate complex splicing patterns, as shown in *Drosophila melanogaster (8, 9)* and human *(10)*.

All these examples indicate that the secondary structure adopted by the pre-mRNA modulates splicing. However, this may be a transient process, since RNA folds co-transcriptionally and the structure may change as more RNA gets produced *(11, 12)*. Furthermore, these structures can be altered by temperature, transcription, or other factors that prevent their formation or stabilize them *(2, 7, 13)*, thus providing more possibilities for splicing regulation. It is still unclear to which extent secondary structures play a role in splicing in human and in general, in metazoans. However, studies in single cell eukaryotes have provided some insights. In contrast to what happens in higher eukaryotes, yeast species do not have as many of the splicing auxiliary factors *(14, 15)*, which reduces dramatically the number of regulatory mechanisms and makes splicing more dependent on *cis* acting elements.

Recent works have suggested that RNA structures could be a general mechanism to explain

3'ss selection in yeast *(16, 17)*, expanding previous observations suggesting a role of RNA structures in splicing regulation in yeast *(18-22)*. This proposed mechanism could resolve, in particular, those cases where a scanning mechanism from the BS onwards *(23)* could not explain splice site selection. Furthermore, secondary structures have been shown to explain some cases of alternative splicing in yeast, in which changes in temperature affect the stability of the RNA structure and thus, produces altered splicing patterns *(17, 24)*.

In this review, we provide the resources and steps to obtain information on the secondary structure of the RNA in relation to splicing, which may serve as starting point for an integrative analysis with multiple other features, for instance, using Machine Learning methodologies *(24)*. In particular, we describe how to calculate optimal and suboptimal secondary structures, how to calculate the effective distance and the accessibility, and how to predict conserved secondary structures affecting splicing. As a practical example, we use the case of the RNA secondary structure in introns that has been shown to be relevant for 3'ss selection in yeast and that could be a general splicing regulatory mechanism *(16, 17, 24)*.

## 2. Materials

In this review we will describe the use of several online tools and databases to retrieve and analyze data. Furthermore, we will illustrate the use of some available programs and simple Perl programs on a unix terminal to perform data analysis such as prediction of RNA structures, calculation of effective distance, and prediction of accessibility. Therefore, a computer with a Unix terminal and Perl programming language installed is required. Other websites and resources used in this review are listed below:

**2.1 Databases and software**

Saccharomyces genome database: http://www.yeastgenome.org/

Ensembl database: www.ensembl.org/

UCSC genome browser: http://genome.ucsc.edu/

Galaxy: https://galaxyproject.org/

Sequence format converter: http://www.ebi.ac.uk/Tools/sfc/

Vienna RNA package: http://www.tbi.univie.ac.at/RNA/

Perl: http://www.perl.org/

## 3. Methods

### 3.1 Retrieving sequence datasets

As splicing often occurs co-transcriptionally *(2)*, we expect that the RNA structures involved in splicing regulation are going to be short and dynamic, i.e. they will not be very stable and may change as the amount of pre-mRNA sequence transcribed increases. Furthermore, we have to consider the scenario in which RNA structures compete with RNA Binding Proteins (RBPs) or small nuclear ribonucleoproteins (snRNPs). Therefore, to predict secondary structure that may affect splicing we will use short sequences around splicing signals (or any other elements of interest such SR protein binding sites). Accordingly, we will need to have some prior knowledge about RBPs or snRNPs that may be involved in the process to limit the amount of sequence to be used. For instance, in the example proposed here, we will use pre-mRNA sequences spanning from the BS to the region downstream of the 3'ss.

The sequence of introns and exons from *S. cerevisiae* can be obtained from several online resources such as Saccharomyces Genome Database (http://www.yeastgenome.org/) *(25)*, Ensembl (www.ensembl.org/) *(26)*, UCSC (http://genome.ucsc.edu/) *(27)* or Galaxy (https://galaxyproject.org/) *(28)*. These resources provide tools to facilitate sequence retrieval for genes and genomic regions; hence, we will not go over this process. As an example we will use the gene SCN1 from yeast. In Figure 1 you can see the sequence of SCN1 pre-mRNA, with the exons in lower case and the intron in upper case. The sequence of the BS and the 3'ss are highlighted in boldface. We will use this sequence to illustrate the analyses described below.

### 3.2 Secondary structure prediction

RNA structure prediction generally involves the search for configurations of maximum base pairing or of minimum free energy (MFE). As this search entails the exploration of an enormous RNA configuration space, different computation methods propose different strategies to arrive at a result. Besides, these methods must also rely on the availability of correct free energy estimates for the base pairings. There are many methods for RNA structure prediction, e.g.: *mfold (29)*, *RNAsubopt (30)*, *RNAfold (31)*. There are also methods that calculate the secondary structure using information from multiple sequences, either from an alignment or performing the alignment simultaneously to the structure prediction, like *RNAalifold (32)*, *evofold (33)*, *RNAz 2.0 (34)*, or *Locarna (35)*.

1. We will use *RNAfold* (http://rna.tbi.univie.ac.at/) *(31)* to make RNA secondary structure predictions using the command line. To make a simple prediction, first we need to get a sequence in Fasta format. From the SCN1 gene, we extract the sequence between the BS and the 3'ss, discarding the first 8nt downstream from the BS A and the 3'ss sequence (see **Note 1**). We save this sequence in Fasta format as shown in Figure 2. The RNA secondary structure for this sequence can be predicted using the program *RNAfold* (see **Note 2**):

    ```
    RNAfold < seq.fa > rna_struct.txt
    ```

2. As can be seen in Figure 3A, the file `rna_struct.txt` contains the sequence and the MFE structure prediction in bracket notation, labeled as (1) and (2), respectively. Furthermore, we also obtain the energy of the predicted structure, in this case, -8.30 kcal · mol$^{-1}$. This command produces an additional file, `scn1_bs_3ss_ss.ps`, which contains the drawing of the MFE structure predicted (Figure 3B). In this structure, base pairings between the nucleotides are shown as lines connecting nucleotides in different parts of the sequence. Nucleotides that are not in any base pair are shown as loops and bulges.

3. We can obtain further information about the stability of MFE structure by calculating the pair probabilities of the base pairs in the MFE structure. Nucleotide pairs with a high pair probability represent very stable base pairs. In contrast, low pair probabilities suggest that that a particular base pair in the structure is not very likely to occur and thus, in the majority of the cases, it will not happen. We can calculate

the RNA secondary structure and the base pair probabilities of the structure using the option –p:

```
RNAfold –p < seq.fa > rna_struct.txt
```

4. In this case, we obtain another additional file, `scn1_bs_3ss_dp.ps`, which contains the pair probabilities of all possible base pairs. We can use this last file to redraw the predicted secondary structure (Figure 3B), adding the probability of the base pairs in the structure, using the program `relplot.pl` from the Vienna RNA package:

```
relplot.pl –p scn1_bs_3ss_ss.ps scn1_bs_3ss_dp.ps> scn1_bs_3ss_rss.ps
```

The structure displaying the pair probabilities, `scn1_bs_3ss_rss.ps`, is shown in Figure 3C. In this case, the nucleotides in the structure are colored according to their probability in the MFE structure.

### 3.2.1 Suboptimal structure prediction

To do a more accurate analysis of the possible secondary structures, we can calculate suboptimal structures that are similar to the MFE but not as probable. Assuming that the structures involved in splicing regulation are transient and unstable, e.g. by occurring along a short time span during transcription, it is plausible that the effect of the RNA secondary structure on splicing is the effect not from a single optimal structure but also from other suboptimal but nearly identical structures. To assess this possibility, one can predict suboptimal structures whose free energy are close to that of the optimal secondary structure using the program *RNAsubopt (30, 22)*. The relation between the stability of a structure and its probability is given by

$$P(S_i) = \frac{1}{Z} e^{-\frac{\Delta G(S_i)}{RT}}$$

where $\Delta G(S_i)$ is the free energy of the structure $S_i$ for sequence $S$,

$$Z = \sum_{S_k} e^{-\frac{\Delta G(S_k)}{RT}}$$

is the partition function of all possible secondary structures $S_k$ of sequence $S$, $R$ is the physical gas constant and *T* is the temperature. This equation determines that the lower the free energy, the higher its probability. Accordingly, structures with energies close to the MFE can still be highly probable. The method *RNAsubopt* calculates a sample of the possible secondary structure space within a given variation of the MFE. Using these suboptimal structures, one can for instance calculate the distribution of effective distances for each of the introns analyzed. This allows determining the effect of the variability of the secondary structure.

1. In our example, we will generate a random sample of 1000 suboptimal structures drawn with probabilities equal to their Boltzmann weights (`-p 1000`) and whose energy does not vary more than 5% from the MFE structure (`-ep 5`).

```
RNAsubopt -ep 5 -p 1000 < seq.fa > subop_rna_structs.txt
```

In this case, the resulting file, `subop_rna_structs.txt`, contains only the secondary structures predicted in bracket notation.

**3.3 Linking RNA structures to splicing regulation**

The two main mechanisms by which a secondary structure can hinder splicing is by (1) affecting the distance between splicing signals (i.e. the BS and the 3'ss), which will reduce splicing efficiency or by (2) blocking the recognition of splicing signals, i.e. changing splicing signal accessibility *(17)*. These two effects can be measured by calculating the effective distance and the nucleotide accessibility.

**3.3.1 Effective distance**

The effective distance is defined as the linear distance in nucleotides (nt) after removing the secondary structure. More specifically, removing all the bases that are part of a structured region and keeping the 2 bases corresponding to the beginning and the end of the structured region. The simplest way of calculating the effective distance between two signals in the RNA (i.e. the BS and the 3'ss) is to predict the MFE structure and calculate the distance between them after discarding the positions included within the secondary structure. To calculate the effective distance we can use a small program in Perl, effective_distance.pl, which will parse the information contained in the RNA structure predicted in bracket notation and will return the effective distance calculated in nucleotides.

```
perl effective_distance.pl < rna_struct.txt > effective_dist.txt
```

The program effective_distance.pl could look like this:

```perl
#!/usr/bin/perl -w
use strict;

my $effective_length = 0;
my $open = 0;
my $close = 0;

while (<STDIN>) {
    next if ($_=~m/>/ || $_=~m/^[AUGC]/);
    chomp;
    my $effective_length = 0;
    my @line = split;
    my @structure = split (//, $line[0]);
    foreach my $i (0..$#structure){
        if ($structure[$i] eq "." && $open == $close){
            $effective_length++;
        }
        elsif ($structure[$i] eq "("){
            $open++;
        }
        elsif ($structure[$i] eq ")"){
            $close++;
            if ($open == $close){
                $effective_length += 2;
            }
        }
    }
    $effective_length = $effective_length+8+3;
    print $effective_length, "\n";
}
close (STDIN);
```

The output given by this program is a number, which represents the effective distance in nucleotides between the BS and the 3'ss. This number, also considers the 8 nt discarded downstream of the BS A at the beginning of the sequence and the 3 nt of the 3'ss, which should be considered to calculate the effective distance *(17)*.

As before, besides the MFE structure, we can also incorporate suboptimal structures to the calculation of the effective distance. In this case, we can run the program `effective_distance.pl` using the suboptimal structures predicted before with *RNAsubopt*.

```
perl effective_distance.pl < subop_rna_structs.txt > effective_dist_subopt.txt
```

The output file contains the effective distance of each of the 1000 suboptimal structures predicted before. Given that the structures predicted are a weighted sample of all possible structures, we can use this data to calculate the mean effective distance of the 3'ss analyzed. In Figure 4 we see the distribution of effective distances calculated for the suboptimal structures. For comparison, we have colored in red the bar for the effective distance obtained from the MFE structure. We observe that the distribution of effective distances is bimodal. Furthermore, the most frequent effective distance in the suboptimal structures predicted (22 nt) differs from that of the optimal structure (28 nt; red bar). Therefore, using only the MFE structure may result in a wrong estimate of the effective distance.

### 3.3.2 Accessibility of splicing signals

When secondary structures are placed overlapping *cis* elements in the sequence, they can hinder the recognition of these elements by other proteins or RNAs. Therefore, when measuring the ability to recognize a signal in an RNA molecule such as a splice site, we will have to measure its accessibility, i.e. whether the signal will be available to other proteins or will be hidden by an RNA structure.

Even though the MFE structure may be the most frequent, we have already shown that suboptimal structures are important to understand the effect of RNA structures in splicing regulation. The pair probability, defined above, can also be calculated considering the contribution from all possible structures. In this way, we will be able to determine a local effect of all structures on the recognition of a splicing signal. Moreover, the pair probability over all possible structures also allows describing the probability of not being in a base pair, i.e. the accessibility. This accessibility is what will actually give us information about the likelihood that a signal in the RNA is accessible to a protein factor to bind, or on the contrary, is likely to be "hidden" inside a secondary structure.

For our present example, we will include the sequence upstream of the 3'ss till the BS and also some nucleotides downstream of the 3'ss, as they can also be included in secondary structures affecting its recognition. In other cases, such in the case of the 5'ss, we will be interested in selecting the sequence in a different way, as only some nucleotides upstream of and downstream the 5'ss may affect its recognition. The pair probability of a given position can be calculated using the program *RNAfold (31)*. From the result given by *RNAfold*, we will calculate the accessibility of the nucleotides from the 3'ss.

1. From the original Fasta sequence, extract the sequence between the BS and the 3'ss, discarding the first 8 nt downstream from the BS A. In this case, we will include the 3'ss and 5 nt downstream of the 3'ss, as we will want to quantify the probability of the 3'ss being included in different secondary structures. We will save this secondary structure in Fasta format, `seq_ext.fa`, as described above (Figure 5).
2. For each of the sequences, predict the RNA secondary structure with *RNAfold* as described in section 3.2. In this case, we will use the option `–noPS,` which avoids producing the postscript figure of the MFE structure:

```
RNAfold –p –noPS < seq_ext.fa > rna_struct_ext.txt
```

As before, the option `–p` will produce a file called `scn1_ext_dp.ps`, which is a dot plot that contains for each pair of nucleotides in the sequence the probability of them being in a base pair. Graphically, the file shows a matrix. Each position in the matrix is represented by a black square whose size is proportional to the probability that a given pair of nucleotides is in a base pair (Figure 6A). The probability of a pair of nucleotides being in a base pair is also provided inside of the dot plot file in multiple lines, each line of the form (Figure 6B)

```
i    j    sqrt(p)    ubox
```

where `i` and `j` are the nucleotides evaluated, `sqrt(p)` is the square root of the pair probability of the base pair between `i` and `j,` and `ubox` indicates that these are the elements above the diagonal, i.e. representing the pair probabilities from all possible structures. The label `lbox` is used for the matrix elements below the diagonal, which represent the pair probabilities of the optimal structure.

3. We will use another small program, `accessibility.pl`, to parse the information inside the dot plot file and calculate the average accessibility of the 3'ss:

```
perl accessibility.pl < scn1_ext_dp.ps > accessibility.txt
```

the program `accessibility.pl` looks like this:

```perl
#!/usr/bin/perl -w
use strict;

my $seq="";
my @pair_probability;
my $seq_flag = 0;

while (<STDIN>) {
    chomp;
    if ($_=~m/^\/sequence\s+\{/){
       $seq_flag =1;
    }
    elsif ($seq_flag == 1){
       if ($_=~m/^\)\s+\}\s+def/){
          $seq_flag=0;
          @pair_probability = split (//,0 x length ($seq));
       }
       else{
          $seq .= $_;
          $seq =~s/\\//g;
       }
    }
    elsif ($_=~m/(\d+)\s+(\d+)\s+([0-9.Ee-]+)\s+ubox/){
       my ($i, $j, $probability) = ($1, $2, $3);
       $probability *=$probability;
       $pair_probability[$i] += $probability;
       $pair_probability[$j] += $probability;
    }
}
close (STDIN);

my @ss = splice (@pair_probability,-8,3);
my $average_pp = ($ss[0]+$ss[1]+$ss [2])/3;
my $average_accessibility = 1 - $average_pp;

print $average_accessibility, "\n";
```

This will return the average accessibility of the 3'ss of interest, which will be saved in the file `accessibility.txt`.

If we want to use the accessibility of a signal to understand if a 3'ss could be functional or not, what we can do is to compare the accessibility of a candidate 3'ss to that of the annotated 3'ss. If we find any candidate 3'ss that have an accessibility similar or higher than a nearby annotated 3'ss and it is in range, i.e. the effective distance between the BS and the 3'ss is not too big, this candidate could be a possible alternative 3'ss. Furthermore, we can also compute the accessibility using sequences of different length, which allows estimating the fact that splicing and transcription are coupled.

**3.4 Conserved secondary structures**

Another aspect in which we can be interested is in the identification of conserved secondary structures, which may be indicative of a mechanism conserved across different species. In human, it has been shown that conserved secondary structures overlapping a splice site are more frequent in alternative exons than in constitutive ones *(5)*, suggesting that structure could actually be a mechanism of splicing regulation conserved across eukaryotes. In this case, we will do an RNA prediction based on a sequence alignment. This prediction can be done with programs such as *RNAalifold (32)* or *evofold (33)*, to which we will have to input an alignment in Clustal format (see **Note 3**) to make the prediction.

1. First, we get the homologous sequences to the one used before to make the prediction. If we know the genomic coordinates of our sequence (in this case, ChrI:87447-87500) we can extract the homologous region from the genomic alignments in UCSC using Galaxy *(28)* (for more details on how to perform this, see the available information at. https://galaxyproject.org/)
2. Using Galaxy we can convert the original alignment format from MAF to Fasta using the *Convert Formats* tool. Additionally, the resulting file, `yeast_all.fa`, should be converted into Clustal format, `yeasts_all.aln`, which can be done with tools like the *Sequence format converter* (http://www.ebi.ac.uk/Tools/sfc/) *(36)*.

3. For each of the sequences, we predict the RNA secondary structure with *RNAalifold*.

   ```
   RNAalifold < yeasts_all.aln > yeast_all.txt
   ```

   As before, we can use the output file of the prediction, `yeast_all.txt`, to calculate the effective distance between the BS and the 3'ss using the program `effective_distance.pl`

4. If we run *RNAalifold* with the option –p and include the 3'ss sequence plus 5nt downstream (as done before), we will produce a file called `alidot.ps` that could be used to measure the accessibility of the 3'ss according to the conserved secondary structure.

**3.5 Significance of results**

In general, the longer the sequence and the higher its GC content, the more likely it is to predict a secondary structure computationally. Accordingly, we must evaluate the significance of our analyses taking into consideration these and other possible biases. One of the most effective ways to assess significance is consider a control set, which would represent the null hypothesis. For the analysis of secondary structures, we can generally consider two types of control sets: randomized sequences and a negative control set. Randomized sequences are obtained from the original set by shuffling nucleotides. Within intron regions, shuffling single nucleotides could be enough, but shuffling while keeping di-nucleotide frequencies can help controlling for more subtle structural biases. For exonic regions, the nucleotide shuffling should be done such that the encoded amino acid sequence, codon usage and base composition of the RNA are preserved *(37)*. By construction, this control set maintains the sequence content and length distribution. On the other hand, when performing an analysis using a multiple alignment, we can consider a different form of shuffling: vertical shuffling. In this method, each column of the alignment is shuffled vertically. In this way, the sequence conservation is preserved, but possible structural dependencies within each sequence are broken. This can also be extended to di- or tri- nucleotides (see *(38)* for an example).

A control set can also be built by extracting random genomic regions that resemble the regions being analyzed, but that are known to be non functional to some extent. For instance, a control set for exons could consist of intronic regions flanked by motifs similar to splice-sites, but have no evidence of being expressed (see *(39)* for an example). Likewise, a control set for intronic regions could be extracted from random intergenic regions of the same sizes, known to be devoid of any expression evidence and selected such that they have a similar sequence content bias (see *(17)* for an example). Significance is then assessed by performing the structure prediction analysis on the control set, exactly in the same way as we did before on our input data set. Properties from both sets, e.g. effective distance, accessibility, frequency for structures per length, can then be compared to obtain a measure of significance, for instance, by using false discovery rate or any other statistical test *(40)*.

## 3. Notes

***Note 1*:** We discard these nucleotides downstream of the BS as it has been shown that they are not generally included in a secondary structure *(17)*.

***Note 2*:** We describe here how to use the programs *RNAfold*, *RNAsubopt* and *RNAalifold* from the command line. However, these and other programs from the Vienna package can also be executed online ( http://www.tbi.univie.ac.at/RNA/ ).

***Note 3***: A file in CLUSTAL format is a plain text file with a header starting with the Word "CLUSTAL" followed by information of the version. Multiple alignment programs generate alignments in this format, possibly adding extra information. The alignment is generally represented in blocks of 60 residues, where each block starts with a sequence identifier. Additionally, the end of each line may include the number of residues in that line of the alignment. Below each block, the symbol "*" indicates whether the position in the alignment is identical for all sequences (see http://www.clustal.org/ for more details). In the case of amino acid alignments, the symbols ":" and "." indicate conserved or semi-conserved substitutions. Below, we show the example of the multiple sequence alignment used for the prediction of the conserved secondary structure using *RNAalifold* (Figure 7).

```
>SNC1(YAL030W)
augucgucaucuacucccuuugacccuuaugcucuauccgagcacgaugaagaacgaccc
cagaauguacagucuaagucaaggacugcggaacuacaagcuGUAAGUACAGAAAGCCAC
AGAGUACCAUCUAGGAAAUUAACAUUA**UACUAACUU**UCUACAUCGUUGAUACUUAUGCGU
AUACAUUCAUAUACGUUCUUCGUGUUUAUUUU**UAG**gaaauugaugauaccgugggaauaa
ugagagauaacauaaauaaaguagcagaaagaggugaaagauuaacguccauugaagaua
aagccgauaaccuagcggucucagcccaaggcuuuaagaggggugccaauagggucagaa
aagccaugugguacaaggaucuaaaaaugaagaugugucuggcuuuaguaaucaucauau
ugcuuguuguaaucaucgucccauugcuguucacuuuagucgauag
```

**Figure 1.** Sequence of the SNC1 gene in Fasta format. Fasta format consist of a header line starting with ">" and additional lines with the sequence data, generally split in blocks of 60 residues. In the figure, the exon sequence is shown in grey lower case letters whereas the intron sequence is shown in black upper case letters. The branch site (BS) sequence (UACUAACUU) and the 3'ss (UAG) are highlighted in bold with the BS A colored in red.

```
>scn1_bs_3ss
AUCGUUGAUACUUAUGCGUAUACAUUCAUAUACGUUCUUCGUGUUUAUUUU
```

**Figure 2.** Intronic sequence between the BS and the 3'ss, discarding the 3'ss sequence and 8nt downstream of the BS A.

```
    >scn1_bs_3ss
(1) AUCGUUGAUACUUAUGCGUAUACAUUCAUAUACGUUCUUCGUGUUUAUUUUU
(2) ......(((((....((((((......))))))....)))))........ ( -8.30)
```

**A**

**B** 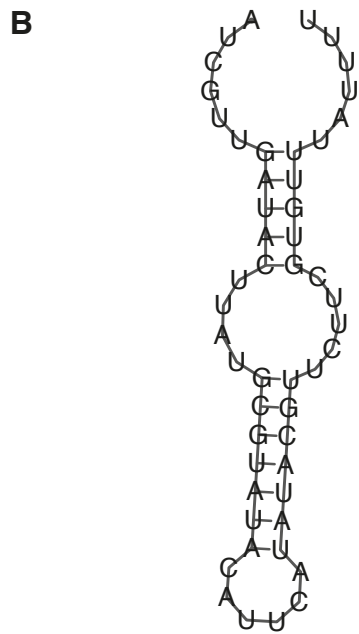

**C** 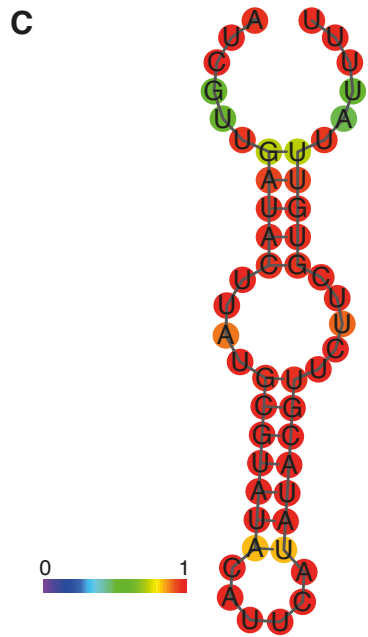

**Figure 3. (A)** MFE structure prediction output by RNAfold. In the output we get (1) the nucleotide sequence given as input and (2) the MFE secondary structure prediction in bracket notation. In this format, "(" and ")" denote positions that are forming a base pair whereas "." correspond to unpaired nucleotides. The energy of the structure, expressed in kcal · mol$^{-1}$ is provided between brackets. **(B)** Graphical representation of the predicted MFE structure **(C)** Graphical representation of the MFE structure showing the pair probabilities of the nucleotides in the MFE structure. For nucleotides outside the secondary structure (i.e. in bulges, loops or unstructured), the color represents the probability of not being in a base pair for the MFE structure in the same scale. The color scale goes from purple, which represent the lowest pair probability to red, which represents the highest probability.

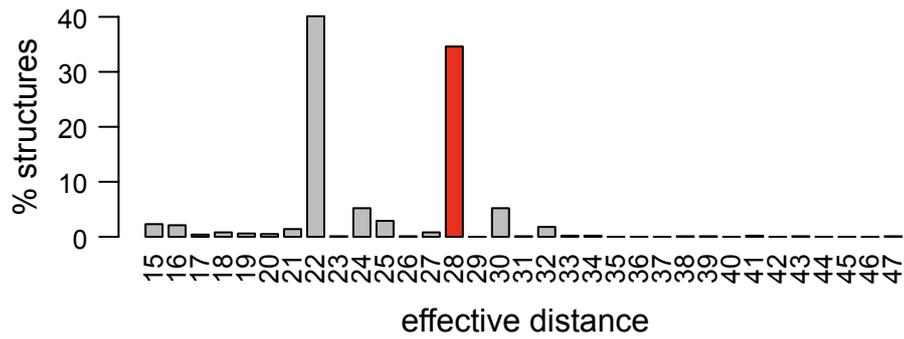

**Figure 4.** Barplot showing the distribution of effective distances (in nucleotides) for the 1000 suboptimal structures predicted. The x-axis shows the effective distances measured in nucleotides. The y-axis shows the % of structures with a given effective distance. The value corresponding to the effective distance of the MFE is indicated with a red bar.

```
>scn1_ext
AUCGUUGAUACUUAUGCGUAUACAUUCAUAUACGUUCUUCGUGUUUAUUUU**UAG**GAAAU
```

**Figure 5.** Intronic sequence between the BS and 5nt downstream of the 3'ss, discarding the 8nt downstream of the BS A. The 3'ss sequence is shown in bold.

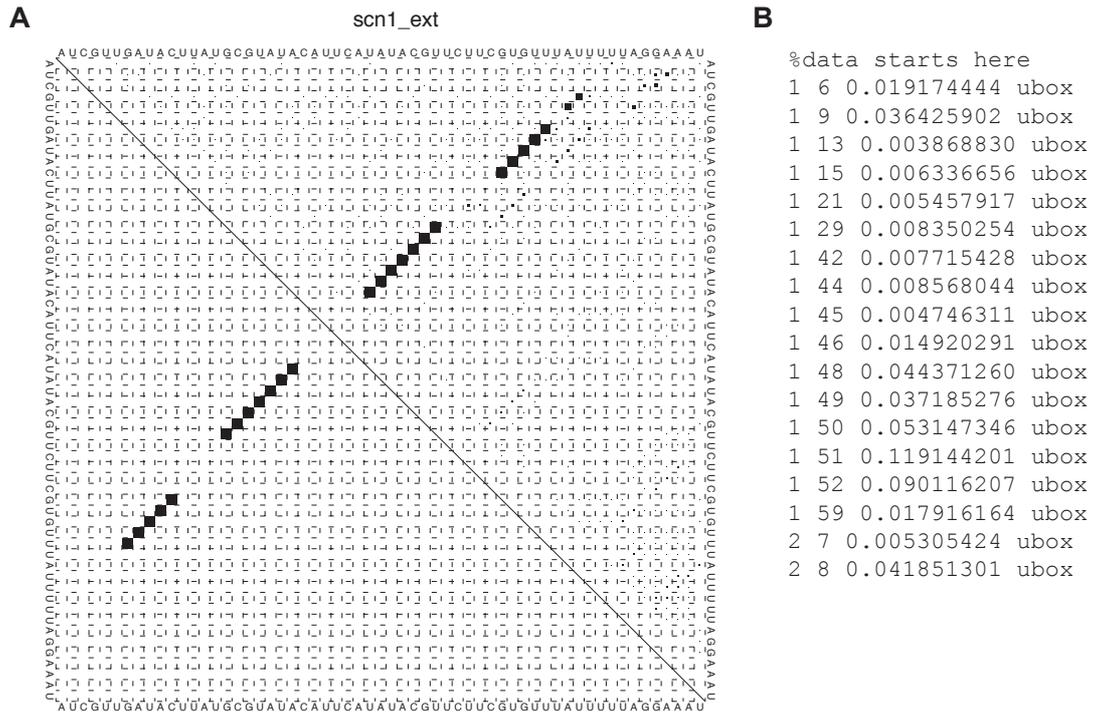

**Figure 6. (A)** Dot plot showing the base pair probabilities. The input sequence is shown at both sides of the matrix. For each pair of nucleotides, *i* and *j*, we have a black square whose size is proportional to the probability of *i* and *j* being in a base pair. The elements above the diagonal (ubox) represent the base pair probabilities calculated from all structures for each pair *i* and *j*. The elements below the diagonal represent the base pair probabilities in the MFE structure for each pair *i* and *j*. Only probabilities larger than $10^{-6}$ are shown. **(B)** The dot plot postscript file also includes the probability of each pair of nucleotides *i j* to be in a base pair in the form: *i, j, square root of the probability, ubox*.

```
CLUSTAL X (1.81) multiple sequence alignment

sacCer3    ATCGTTGATACTTATGCGTATAC-ATTCATATACG-TTCTTCGTGTTTAT-TTTTAG
sacPar     GTCATTGATATATATACGTATAC-ATACGTGTACG-TATGCCGTGTTTAT-TTTTAG
sacMik     GTCGTTAATGTTTTTACGTATAT-GTATGTATACG-TATATCACGTTATT-TTACAG
sacKud     GACATTGATGTACATACGCATACGGTGTATGTACATTTTTTCATGTTTTTCTTCCAG
sacBay     GACATTACTGTATATACGTATAC-GTTTATGTATG-T------CGTTATCTTCATAG
sacKlu     ----------------------------------------------TTTTT-TAACAG
                                                          **    *   **
```

**Figure 7.** Nucleotide sequence alignment in Clustal format. The alignment has been extracted from the 7-way genome alignment from UCSC for yeast species, for the region between the BS and the 3'ss (excluding the BS signal). The species included in the alignment are *S. cerevisiae* (sacCer3), *S. paradodux* (sacPar), *S. mikatae* (sacMik), *S. kudriavzevii* (sacKud), *S. bayanus* (sacBay) and *S. Kluyveri* (sacKlu).